\documentclass[conference]{IEEEtran}
\IEEEoverridecommandlockouts
\usepackage{cite}
\usepackage{amsmath,amssymb,amsfonts}
\usepackage{algorithmic}
\usepackage{graphicx}
\usepackage{textcomp}
\usepackage{xcolor}
\usepackage{amsmath,graphicx}
\usepackage{algorithm}  
\usepackage{algorithmic}
\usepackage{multirow}
\usepackage{bigstrut}
\usepackage{amsmath}
\usepackage{cite}
\usepackage{booktabs}
\usepackage{array}
\def\BibTeX{{\rm B\kern-.05em{\sc i\kern-.025em b}\kern-.08em
    T\kern-.1667em\lower.7ex\hbox{E}\kern-.125emX}}
\begin{document}

\title{An Optimized H.266/VVC Software Decoder On Mobile Platform}

\author{\IEEEauthorblockN{Yiming Li, Shan Liu, Yu Chen, Yushan Zheng, Sijia Chen, Bin Zhu, Jian Lou}
	\IEEEauthorblockA{{Tencent Media Lab, Shenzhen, China and Palo Alto, CA, USA,} \\
		{\{marcli, shanl\}@tencent.com}}
                        
}

\maketitle

\begin{abstract}
As the successor of H.265/HEVC, the new versatile video coding standard (H.266/VVC) can provide up to 50\% bitrate saving with the same subjective quality, at the cost of increased decoding complexity. To accelerate the application of the new coding standard, a real-time H.266/VVC software decoder that can support various platforms is implemented, where SIMD technologies, parallelism optimization, and the acceleration strategies based on the characteristics of each coding tool are applied. As the mobile devices have become an essential carrier for video services nowadays, the mentioned optimization efforts are not only implemented for the x86 platform, but more importantly utilized to highly optimize the decoding performance on the ARM platform in this work. The experimental results show that when running on the Apple A14 SoC (iPhone 12pro), the average single-thread decoding speed of the present implementation can achieve 53fps (RA and LB) for full HD (1080p) bitstreams generated by VTM-11.0 reference software using 8bit Common Test Conditions (CTC). When multi-threading is enabled, an average of 32 fps (RA) can be achieved when decoding the 4K bitstreams.
\end{abstract}

\begin{IEEEkeywords}
H.266, Versatile Video Coding (VVC), Video decoding, ARM, SIMD optimization
\end{IEEEkeywords}

\section{Introduction}
With the popularization of the video applications on the Internet and the demand for high-quality video services, a more efficient video compression standard with high coding performance is desired. After High Efficiency Video Coding (HEVC)/H.265 standard, Joint Video Experts Team (JVET) between VCEG (Q6/16) and ISO/IEC JTC1 SC29/WG11 (MPEG) was created, and formed the Versatile Video Coding standard (H.266/VVC) in July 2020 \cite{b1}. It can provide up to 50\% improvement in compression efficiency for the same subjective quality compared to its predecessor HEVC \cite{b2}. In order to enhance the coding performance, several coding tools with decoder side computations are adopted, like decoder side motion vector refinement (DMVR), bi-directional optical flow (BDOF), adaptive loop filter (ALF), etc. These coding tools introduce additional complexities for the decoder, that makes it a hard work to optimize the decoding speed.

In general, video decoders can be classified into hardware implementations and software implementations, where hardware decoders are preferred in low-power devices, but normally come a few years later after a standard is finalized. In contrast to the hardware decoder, software decoders may need more power, but still play an important role on general computer, especially in the early stage for the new coding standard. Therefore, it is essential to have an efficient and optimized software decoder implementation to support the emerging applications. In \cite{b3}\cite{b4}, an independent VVC software decoder implemented by Tencent demonstrated real-time HD/UHD decoding capability on x86 platform.  Considering that mobile devices have become an essential carrier and display tool for video services, extensive optimization efforts were made on top of the framework of \cite{b3} to achieve real-time HD/UHD decoding on the mobile platform. As a result, a uniform-designed software H.266/VVC decoder that can run real-time on different platforms and supports versatile functionalities such as screen content coding (SCC) is accomplished. This paper will focus more on the discussions of mobile optimization and capabilities. The rest of the paper is organized as follows: Section \ref{sec2} presents an overview of H.266/VVC decoder processing blocks and briefly introduces the proposed software decoder design including single instruction, multiple data (SIMD) processing technology and data/task level parallelism. Section \ref{sec3} elaborates the optimization on some important modules and shows the benefit of the implementation. Experimental results of the proposed decoder are shown in Section \ref{sec4}. Finally, Section \ref{sec5} concludes the paper.

\section{Coding Tool Independent Decoder Optimization}\label{sec2}

\subsection{Overview of H.266/VVC Decoder}\label{sec2.1}

\begin{figure}[htbp]
	\centerline{\includegraphics[width=8cm]{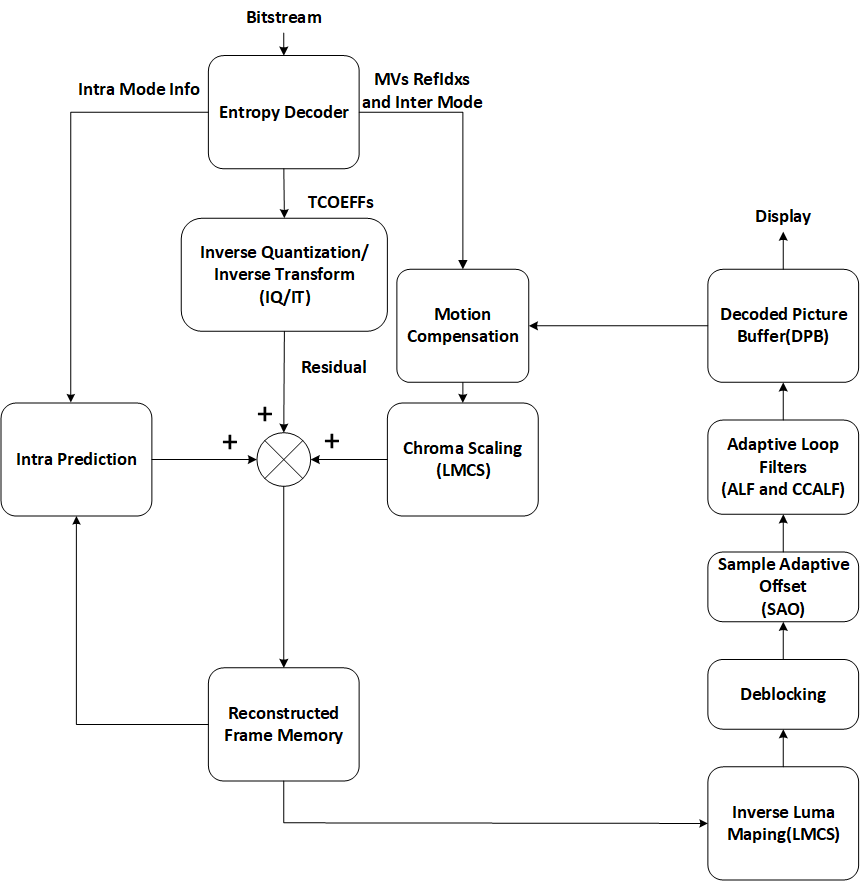}}
	\caption{VVC Decoder Structure \cite{b3}.}
	\label{fig1}
\end{figure}

The structure of H.266/VVC decoder with the main processing modules is shown in Fig. \ref{fig1}. Following the decoding pipeline, the entropy decoder stage is the first stage, which extract and decode the codewords from the input bitstream. After that, the syntax elements are derived. Similar as H.265/HEVC, the entropy encoding/decoding is still based on the context-adaptive binary arithmetic coding (CABAC) scheme, but several enhancements have been introduced into VVC to improve the throughput and the compression efficiency.

The derived intra mode after the entropy decoder stage is used in intra prediction stage, where the prediction value can be obtained by using the boundary pixels and the intra mode index. To increase the prediction accuracy, VVC extends the intra modes from 35 to 67, and wide-angle intra prediction modes are used for the non-square blocks. Different with the traditional angular prediction, the matrix weighted intra prediction (MIP) technology takes the left and above neighbouring boundary samples as the input to generate the prediction by matrix vector multiplication and linear interpolation. Besides, a cross-component linear model (CCLM) prediction mode is also introduced in the VVC to utilize the correlation between luma and chroma components. Benefiting from these tools, an accurate predicted block is prepared, and the final reconstructed block can be formed by adding the residual block and predicted block.

\setlength{\belowcaptionskip}{-1cm}

Residual block is generated by the inverse quantization and inverse transform (IQ/IT) stage shown in Fig. \ref{fig1}. This stage can dequantize and transform the frequency-domain coefficients back to the spatial domain. In VVC, multiple transform selection (MTS) scheme is adopted, so that besides DCT-2, DCT8/DST7 can also be selected as the core transform matrices. This tool can help to remove the correlation and reduce the coding bits. As another new coding tool, low-frequency non-separable transform (LFNST) is added to further remove the correlation in transform domain.

As videos usually contain temporal redundancy, for eliminating the correlation in temporal domain, inter prediction technology is widely used, where the motion compensation (MC) stage generates the inter prediction from a decoded reference block. For acquiring a more accurate prediction result, the decoded inter modes from entropy decoding stage. e.g., MVs, RefIdxs can help to locate the reference block. As Fig. \ref{fig1} shows, another input for motion compensation stage is the reference block, which is from the decoded picture buffer (DPB). In the MC stage in VVC, 8-tap interpolation filter is used for luma block and 4-tap filter is used for chroma block. Besides the traditional translation motion as H.265/HEVC used, the affine motion model is also applied in VVC. For further increasing the prediction accuracy, BDOF and DMVR can help to refine the subblock motion vector in order to obtain a better prediction. Similar as the intra prediction module, the reconstructed block generated by inter prediction mode is from the residual block and the prediction block.

In VVC, loop filtering (LF) stage contains three kinds of in-loop filters, including deblocking filter (DBK), adaptive loop filter (ALF) and sample-adaptive offset (SAO). The general idea for DBK and SAO in VVC is similar as the previous video coding standard except the granularity for luma component is changed to 4$\times$4 on the deblocking filtering process. Adaptive loop filters is newly added in VVC. Based on the gradients of each block, one among 25 filters is applied for each 4$\times$4 block filtering.

\begin{figure}[t]
	\centerline{\includegraphics[width=7cm]{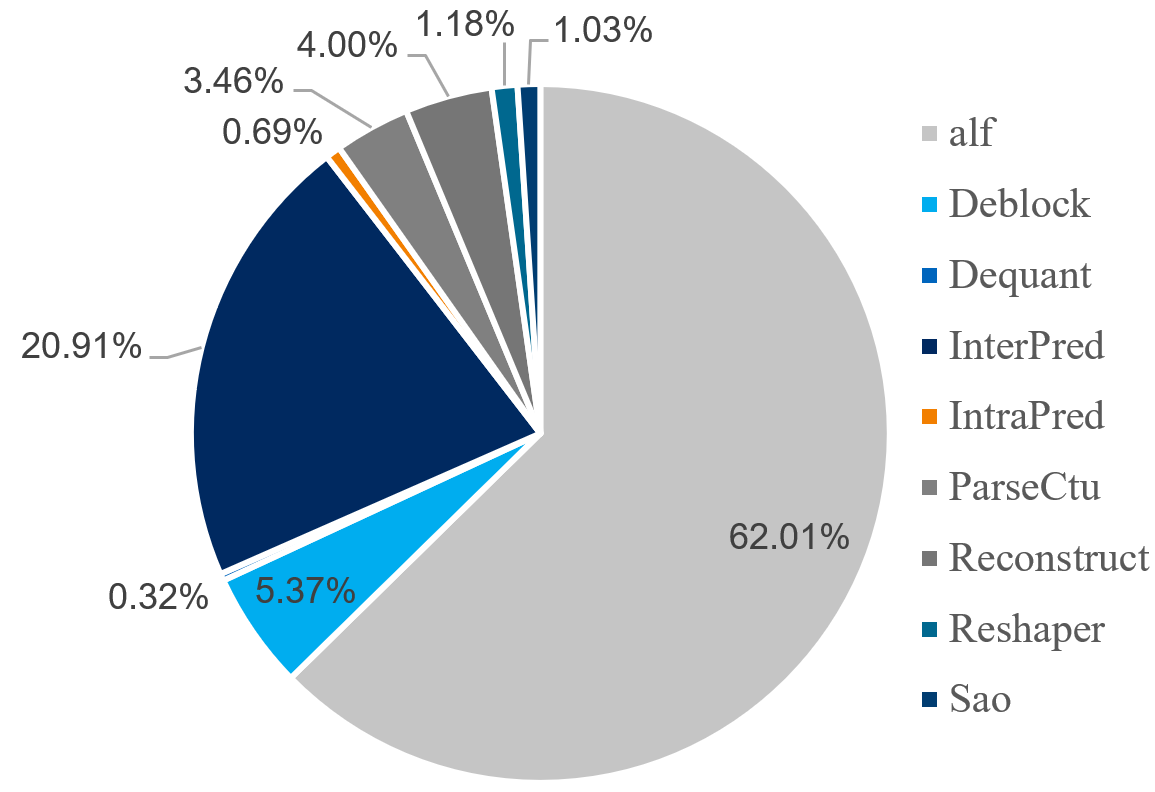}}
	\caption{Runtime Spending of Each Processing Block on ARM Platform without SIMD Acceleration.}
	\label{fig2}
\end{figure}

The test on each new coding tool has been conducted by AHG13 of JVET \cite{b5}, which is based on the reference software. In \cite{b3}, the runtime spent on each processing block based on the optimized real-time decoder is provided on x86 platform. Considering the architecture is different between x86 and ARM platform, the runtime spent of the optimized decoder \cite{b3} is tested on the ARM platform, as shown in Fig. \ref{fig2}. It can be seen as a start point for the optimization on the ARM platform. In this test, the test bitstreams are encoded from the H.266/VVC common test conditions (CTC) \cite{b6} with random access (RA) configuration. We take the average of the 1080p test bitstreams for collecting the runtime in order to reduce the content based fluctuation. As SIMD optimization for x86 platform cannot be directly used in ARM platform, there is no SIMD acceleration in this test. The results show that ALF and inter prediction are the most two time-consuming blocks in the whole decoding pipeline.

\subsection{Tool Independent Decoder Optimization}
As described in Section \ref{sec2.1}, many new coding tools are added in VVC. Besides the coding tools in each module, the new standard can also support larger block size, multi-type coding unit partitions, etc. For versatile video applications, the coding tools for screen content sequences like intra block copy are also included in the main profile of H.266/VVC instead of in the extension \cite{b7}. With the support for the new coding tools and new coding features, the new standard is significantly more complex than previous codecs and more efforts are required for fully accelerating.

\begin{figure}[t]
	\centerline{\includegraphics[width=7cm]{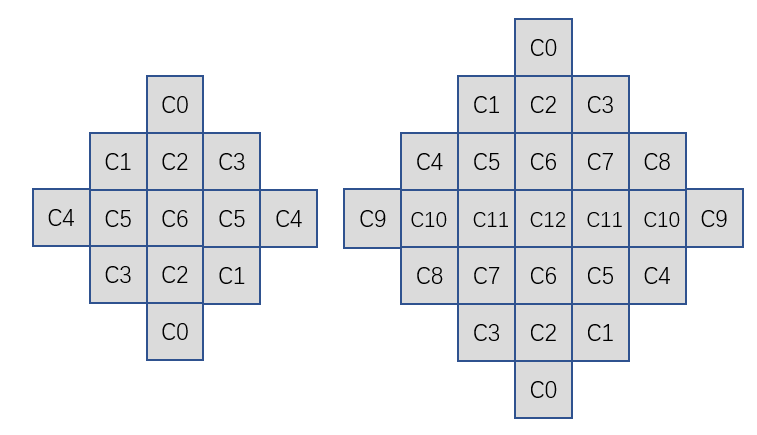}}
	\caption{ALF Filter Shapes (Chroma: 5$\times$5 Diamond, Luma: 7$\times$7 Diamond).}
	\label{fig3}
\end{figure}

SIMD technology operates on loading multiple data in a single operation. If the internal bit depth is 16, and taking 128-bit SIMD, it can load up to eight data points at once and help to accelerate the processing efficiency 8 times when compared with non-SIMD optimized system. On the x86 PC platform, AVX2 instruction sets are widely used, which can support 128-bit and 256-bit parallelism. For the ARM platform, the instruction sets are different, where the ARM NEON intrinsics can only support 64-bit and 128-bit parallelism. Besides, the shuffle operation has many differences on the x86 and ARM platform as AVX2 provide more freedom for the shuffle intrinsics. For designing a decoder that can support different platforms, the SIMD optimization is necessary to have different design in different platforms. On the ARM platform, as the ability for SIMD parallelism is weaker than the x86 platform, some optimization methods is changed according to the characteristics of each tool, which will be described detailed in Section \ref{sec3}. 

Besides the SIMD optimization, the parallel processing for each decoding module can benefit a lot for speedup. In this part, the ARM platform and x86 platform share the same design \cite{b3}, where multiple level parallelization strategies are supported, including picture level, CTU level, task level and sub-CTU level. The picture level parallelism supports decoding multiple nal units (NALU) at the same time when there is no inter dependency. For the CTU level parallelism, CTUs located at different CTU rows are processed following the wavefront pattern at the same time. Task parallelization helps to process multiple tasks of the decoder pipeline without dependency at the same time. And the sub-CTU level parallelization support process multiple CUs within one CTU once the motion vector has been derived.

\section{Key Coding Tool Optimization}\label{sec3}
In order to make the decoder support the ARM platform, the optimization methods based on the characteristic of the coding tools are utilized. In this Section, the optimization strategies on some high-complexity modules are presented.

\subsection{Optimization on ALF}
Adaptive Loop Filter (ALF) is the most time-consuming part on the decoder side. Based on the direction and the activity of local gradients, each 4$\times$4 block will select a set of filter coefficients among 25 filters. As shown in Fig. \ref{fig3}, The shape of filter is diamond, where 7$\times$7 diamond shape is applied for luma component and chroma components use the 5$\times$5 diamond shape. 

Considering the filtering process is very time-consuming, many efforts are spent in this part in the VVC standardization stage. In one 4$\times$4 grid, the luma component may use the rotated filter coefficients, and then applying the filtering process as the 7$\times$7 diamond shape. While for the chroma components, rotation process is unnecessary. The shape length for chroma components are also reduced to 5$\times$5 for complexity reduction. In spite of this, the time cost for calculating the filtered value is still more than half of the whole time in the ALF processing pipeline. Besides using the NEON intrinsics for SIMD optimization directly, the characteristic that the granularity of ALF is considered. First, all of ALF filters and clips values of two horizontal adjacent 4$\times$4 blocks should be obtained at the same time, as the bit length for two horizontal adjacent blocks with block width equals to 4 is exactly 128 bits, which can utilize the 128-bit SIMD efficiently. Second, as NEON only supports shuffle processing in byte level, a possible way to use shuffle processing in word level is found to support filter coefficient rotating process. Third, considering the NEON register in ALF filter is very time consuming, it is preferred to load multiple 4$\times$4 blocks data to do ALF processing with NEON 128 bit registers at the same time, which can obtain a better efficiency.

\vspace{-0.3cm}
\renewcommand\arraystretch{0.9}
\begin{table}[htbp]
	\centering
	\footnotesize
	\caption{Coefficients for 8-Tap Filters}
	\begin{tabular}{|c|c|c|c|c|c|c|c|c|}
		\hline
		\textbf{Fractional sample} & \multicolumn{8}{c|}{\textbf{Interpolation filter coefficients}} \bigstrut[t]\\
		\cline{2-9}     \textbf{position} & f0   & f1   & f2   & f3   & f4   & f5   & f6   & f7 \bigstrut[t]\\
		\hline
		1    & 0    & 1    & -3   & 63   & 4    & -2   & 1    & 0 \bigstrut[t]\\
		\hline
		2    & -1   & 2    & -5   & 62   & 8    & -3   & 1    & 0 \bigstrut[t]\\
		\hline
		3    & -1   & 3    & -8   & 60   & 13   & -4   & 1    & 0 \bigstrut[t]\\
		\hline
		4    & -1   & 4    & -10  & 58   & 17   & -5   & 1    & 0 \bigstrut[t]\\
		\hline
		5    & -1   & 4    & -11  & 52   & 26   & -8   & 3    & -1 \bigstrut[t]\\
		\hline
		6    & -1   & 3    & -9   & 47   & 31   & -10  & 4    & -1 \bigstrut[t]\\
		\hline
		7    & -1   & 4    & -11  & 45   & 34   & -10  & 4    & -1 \bigstrut[t]\\
		\hline
		8    & \multirow{2}[2]{*}{-1} & \multirow{2}[2]{*}{4} & \multirow{2}[2]{*}{-11} & \multirow{2}[2]{*}{40} & \multirow{2}[2]{*}{40} & \multirow{2}[2]{*}{-11} & \multirow{2}[2]{*}{4} & \multirow{2}[2]{*}{-1} \bigstrut[t]\\
		(hpelIfIdx=0) &      &      &      &      &      &      &      &  \bigstrut[t]\\
		\hline
		8    & \multirow{2}[4]{*}{0} & \multirow{2}[4]{*}{3} & \multirow{2}[4]{*}{9} & \multirow{2}[4]{*}{20} & \multirow{2}[4]{*}{20} & \multirow{2}[4]{*}{9} & \multirow{2}[4]{*}{3} & \multirow{2}[4]{*}{0} \bigstrut[t]\\
		(hpelIfIdx=1) &      &      &      &      &      &      &      &  \bigstrut[t]\\
		\hline
		9    & -1   & 4    & -10  & 34   & 45   & -11  & 4    & -1 \bigstrut[t]\\
		\hline
		10   & -1   & 4    & -10  & 31   & 47   & -9   & 3    & -1 \bigstrut[t]\\
		\hline
		11   & -1   & 3    & -8   & 26   & 52   & -11  & 4    & -1 \bigstrut[t]\\
		\hline
		12   & 0    & 1    & -5   & 17   & 58   & -10  & 4    & -1 \bigstrut[t]\\
		\hline
		13   & 0    & 1    & -4   & 13   & 60   & -8   & 3    & -1 \bigstrut[t]\\
		\hline
		14   & 0    & 1    & -3   & 8    & 62   & -5   & 2    & -1 \bigstrut[t]\\
		\hline
		15   & 0    & 1    & -2   & 4    & 63   & -3   & 1    & 0 \bigstrut[t]\\
		\hline
	\end{tabular}%
	\label{table1}%
\end{table}%

\subsection{Optimization on Interpolation Filter}

Interpolation module is also a time-consuming part in H.266/VVC decoder side, which uses filters to help generate a better motion compensated block, where 8-tap filer is used for luma component, 4-tap filter is used for chroma component. In the module of decoder side motion vector refinement (DMVR), 2-tap filter is used. In order to reduce the calculation amount and speedup the interpolation process, the characteristics of the filter coefficients distribution should be taken in consideration. Table \ref{table1} shows the  coefficients for 8-tap filter as an example.

From Table \ref{table1}, it can be seen that only if the sample in the middle fractional position has full 8-tap filter coefficients, for fractional position smaller than 5 or larger than 11, only 7-tap filter is necessary. Thus, we can save several times of the multiplication computation for each pixels in this case. Based on this, the interface of the SIMD function is divided according to the fractional position, which makes different strategies can be designed adapted to the actual filter length for speedup. Considering the symmetric characteristic, fractional sample position equals to 1 and 15 share the same function for code clean, which may help reducing the size of the decoder library. In addition to the separation according to the fractional position, we classify the filter set according to the input and output bit depth as well, as different bit depth should use different function to reduce the logic check times.

\section{Experimental Results}\label{sec4}

\renewcommand\arraystretch{0.9}
\begin{table}
	\centering
	\footnotesize
	\begin{minipage}{0.95\linewidth}
	\centering
	\caption{Information for Test Materials}
	\begin{tabular}{|c|p{4.19em}|c|c|c|}
		\hline
		\multirow{3}[2]{*}{\textbf{Class}} & \multicolumn{1}{c|}{\multirow{3}[2]{*}{\textbf{Sequence}}} & \multicolumn{1}{c|}{\multirow{3}[2]{*}{\textbf{Frames}}} & {\textbf{Bitrate}} & {\textbf{Bitrate}} \bigstrut[t]\\
		& \multicolumn{1}{c|}{} &      & { [Mbps]} & { {[Mbps]}} \\
		& \multicolumn{1}{c|}{} &      & {{/RA}} & {{/LDB}} \bigstrut[t]\\
		\hline
		\multirow{6}[12]{*}{ClassA} & \multicolumn{1}{c|}{Tango} & 294  & 16.57 & {/} \bigstrut[t]\\
		\cline{2-5}         & \multicolumn{1}{c|}{FoodMarket} & 300  & 14.66 & {/} \bigstrut[t]\\
		\cline{2-5}         & \multicolumn{1}{c|}{Campfire} & 300  & 42.33 & {/} \bigstrut[t]\\
		\cline{2-5}         & \multicolumn{1}{c|}{CatRobot} & 300  & 17.45 & {/} \bigstrut[t]\\
		\cline{2-5}         & \multicolumn{1}{c|}{DaylightRoad} & 300  & 22.73 & {/} \bigstrut[t]\\
		\cline{2-5}         & \multicolumn{1}{c|}{ParkRunning} & 300  & 95.98 & {/} \bigstrut[t]\\
		\hline
		\multicolumn{1}{|c|}{\multirow{5}[10]{*}{ClassB}} & MarketPlace & 600  & 11.49 & 14.96 \bigstrut[t]\\
		\cline{2-5}         & RitualDance & 600  & 8.96 & 9.92 \bigstrut[t]\\
		\cline{2-5}         & Cactus & 500  & 8.98 & 14.53 \bigstrut[t]\\
		\cline{2-5}         & BasketballDrive & 500  & 10.39 & 15.14 \bigstrut[t]\\
		\cline{2-5}         & BQTerrace & 600  & 17.90 & 37.98 \bigstrut[t]\\
		\hline
		\multicolumn{1}{|c|}{\multirow{5}[10]{*}{ClassSCC-1080p}} & ArenaOfValor & 300  & 11.66 & 12.29 \bigstrut[t]\\
		\cline{2-5}         & FlyingGraphics & 600  & 23.10 & 25.70 \bigstrut[t]\\
		\cline{2-5}         & Desktop & 600  & 1.98 & 1.09 \bigstrut[t]\\
		\cline{2-5}         & Console & 600  & 3.64 & 3.29 \bigstrut[t]\\
		\cline{2-5}         & ChineseEditing & 600  & 3.89 & 2.33 \bigstrut[t]\\
		\hline
	\end{tabular}%
	\label{table2}%
\end{minipage}
\end{table}

\renewcommand\arraystretch{0.9}
\begin{table}
	\centering
	\footnotesize

	\begin{minipage}{0.95\linewidth}
		\centering
		\footnotesize
		\caption{The Performance on iOS/Android Platforms}
		\begin{tabular}{|p{4em}<{\centering}|p{5em}<{\centering}|p{2.7em}<{\centering}|p{2.7em}<{\centering}|p{2.7em}<{\centering}|p{2.7em}<{\centering}|}
			\hline
			\multirow{3}[2]{*}{} & \multirow{3}[2]{*}{\textbf{Class}} & \textbf{VTM}  & \textbf{O266}  & \textbf{O266} & \textbf{O266} \bigstrut[t]\\
			&      &  Thread 1 & Thread 1 &  Thread 2 &  Thread Full \\
			&      &   [fps] &  [fps] &  [fps] &  [fps] \bigstrut[t]\\
			\hline
			& ClassA-RA & 2    & 11   & 19   & 32 \bigstrut[t]\\
			\cline{2-6}    iOS  & ClassB-RA & 9    & 53   & 86   & 129 \bigstrut[t]\\
			\cline{2-6}    (A14) & ClassB-LB & 8    & 50   & 80   & 102 \bigstrut[t]\\
			\cline{2-6}    6core & SCC-RA  & 13   & 82   & 126  & 180 \bigstrut[t]\\
			\cline{2-6}         & SCC-LB  & 14   & 88   & 145  & 168 \bigstrut[t]\\
			\hline
			Android & ClassA-RA & /    & 6    & 10   & 20 \bigstrut[t]\\
			\cline{2-6}    (Snap- & ClassB-RA & /    & 26   & 38   & 75 \bigstrut[t]\\
			\cline{2-6}    dragon- & ClassB-LB & /    & 26   & 40   & 59 \bigstrut[t]\\
			\cline{2-6}    865) & SCC-RA & /    & 41   & 60   & 104 \bigstrut[t]\\
			\cline{2-6}    8core    & SCC-LB & /    & 43   & 63   & 85 \bigstrut[t]\\
			\hline
		\end{tabular}%
		\label{table3}%
	\end{minipage}
	
	\begin{minipage}{0.95\linewidth}
		\centering
		\footnotesize
		\caption{The Speedup Ratio for Multi-Threading and SIMD}
\begin{tabular}{|c|c|c|c|c|c|}
	\hline
	\multirow{3}[1]{*}{Class} & Thread & SIMD & O266 & T-2  & T-6  \bigstrut[t]\\
	& 1 w/o &  Speed- &  Thread  & Speedup & \multicolumn{1}{p{3.5em}|}{Speedup} \\
	& NEON & up Ratio & 1[fps] & vs T-1 & \multicolumn{1}{p{3.5em}|}{vs T-1} \\
	\hline
	ClassA-RA & 3    & 3.53 & 11   & 1.69 & 2.87 \bigstrut[t]\\
	\hline
	ClassB-RA & 14   & 3.78 & 53   & 1.62 & 2.44 \bigstrut[t]\\
	\hline
	ClassB-LB & 13   & 3.73 & 50   & 1.59 & 2.03 \bigstrut[t]\\
	\hline
	SCC-RA & 26   & 3.08 & 82   & 1.54 & 2.21 \bigstrut[t]\\
	\hline
	SCC-LB & 28   & 3.15 & 88   & 1.63 & 1.90 \bigstrut[t]\\
	\hline
	Average & /    & 3.45 & /    & 1.61 & 2.29 \bigstrut[t]\\
	\hline
\end{tabular}%
		\label{table4}%
	\end{minipage}

\end{table}

To verify the decoding performance, the results and analysis for the optimized software decoder on different mobile platforms are provided. Test bitstreams are generated by the VTM11.0 reference software under the common test condition (CTC) \cite{b6} with HD and UHD YUV-420 test sequences and 8-bit internal bit-depth. The information for the test materials is shown in Table \ref{table2}, where the QP value for encoding each test sequence is in the range from 22 to 37 following the CTC. The maximum bitrate for the test sequences under random access (RA) configuration and low delay B (LDB) configuration are provided, which can provide the impressions of the decoding complexity for each sequence.

For showing the decoding performance on different platforms, two high-end smartphones with the top SoC on iOS and Android platforms are used in this test, which is iPhone 12pro with Apple A14 SoC and Vivo IQOO3 with Snapdragon 865 SoC. Table \ref{table3} shows the performance on these two different platforms. For A14 SoC, the maximum core number is 6, while for Snapdragon 865, the core number is 8. Thus the label “Thread-full” in Table \ref{table3} means 6 or 8 separately. In this Table, the proposed real-time software H.266/VVC decoder is marked as O266, and the reference software is VTM, where we can see the 4K stream can achieve above 30fps on A14 platform when using 6-threads parallelism. If the single thread mode is used, 1080p bitstream can be decoded in real-time based on A14 SoC. When compared with VTM, the acceleration ratio is about 5.5 times for different classes. In addition, from the experimental results, as the screen content sequences can be predicted well by the intra block copy technology, the bitrate is smaller than the test sequences with natural scene, which makes the decoding speed for screen content test sequences are higher than the common videos. In this test, the decoding speed for screen content test sequences can achieve averagely up to 180 fps on iPhone 12 pro. As the performance of Snapdragon 865 is lower than the A14 SoC, the decoding speed on the Android platform is smaller than the iOS platform, where the average decoding speed for class B is averagely 26 fps on single thread and can achieve up to 75 fps on 8 threads. For screen content sequences, Snapdragon 865 can decode them in real-time, which is higher than 30fps on single thread, and has up to 104 fps on 8 threads. 

\begin{figure}[t]
	\centerline{\includegraphics[width=7cm]{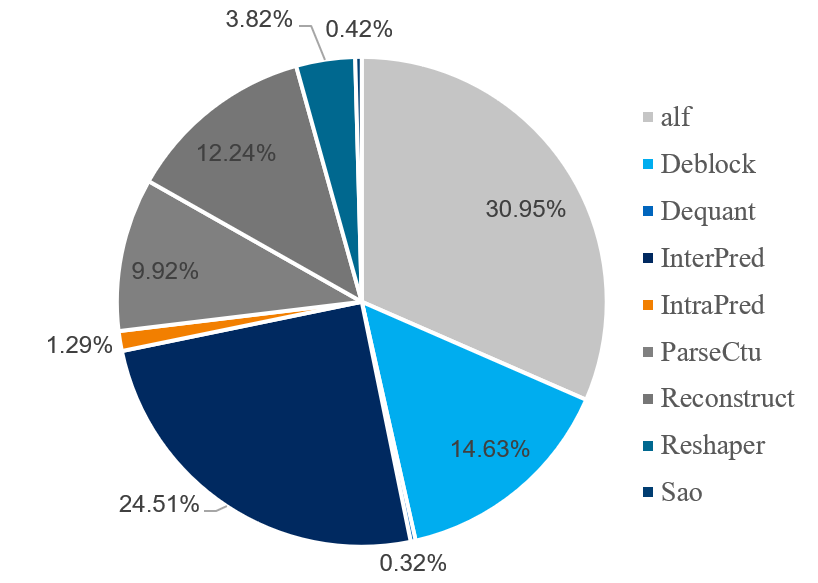}}
	\caption{Runtime Spending of Each Processing Block on ARM Platform with SIMD Acceleration.}
	\label{fig4}
\end{figure}

Based on the analysis of Table \ref{table3}, the speedup ratio for multi-threading can be obtained, which is summarized in the Table \ref{table4}. When compared with the speed of single-thread decoding and 2-threads decoding on iPhone 12pro, the average speedup ratio is about 1.61, while for 6-threads parallelism, the speedup ratio is 2.29. The non-linear speedup ratio increase is due to the chip design of A14 SoC, as there are only two high-performance processors among the 6-core SoC. In this table, the speed up ratio for SIMD acceleration is also provided, where the average speedup ratio is 3.45. Noted, because different modules have different internal bit depth in the calculation process, the speedup ratio for each module is different. Fig. \ref{fig4} shows the runtime ratio information of each processing block on ARM platform with SIMD acceleration. As ALF is the most time-consuming part, and the design is appropriate to obtain high speedup ratio from NEON intrinsics, many efforts are put in this part for SIMD acceleration. When compared with the runtime ratio of each module without SIMD in Fig. \ref{fig2}, the proportion of ALF module is decreased from 62.01\% to 30.95\%, which indicates the speedup ratio of ALF benefiting from SIMD is much higher than other modules.

Since the Apple A14 and Snapdragon 865 SoC are both the processors which have the best quality on their own platform, we also conduct some tests on the iPhone 11pro with A13 SoC and iPhone Xs Max with A12 SoC to obtain the performance on different devices. The experimental results are shown in Table \ref{table5}, where the decoding speed on A12 SoC is slightly higher than 30fps on class B test sequences. Thus, for real applications on iOS platform, it is suggested to use the devices with the performance equal or higher than Apple A12 SoC and enable multi-threading decoding to prevent the performance drop when decoding the high complexity bitstreams. In addition, the decoding speed of  the per-sequence results for the Class B test sequences under RA configuration is also provided in Table \ref{table6} to show more detailed information.

\renewcommand\arraystretch{0.9}
\begin{table}
	\centering
	\footnotesize

	\begin{minipage}{0.95\linewidth}
		\centering
		\footnotesize
		\caption{The Performance on Different iOS Devices}
\begin{tabular}{|c|c|c|c|c|}
	\hline
	\multirow{3}[1]{*}{Class} & VTM   & O266  & O266  & O266  \bigstrut[t]\\
	&  [fps] &  Thread1  &  Thread1  &  Thread1  \\
	&  (A14) & [fps] (A14) & [fps] (A13) & [fps] (A12) \\
	\hline
	ClassA-RA & 2    & 11   & 9    & 7 \bigstrut[t]\\
	\hline
	ClassB-RA & 9    & 53   & 41   & 32 \bigstrut[t]\\
	\hline
	ClassB-LB & 8    & 50   & 42   & 32 \bigstrut[t]\\
	\hline
	SCC-RA  & 13   & 82   & 60   & 49 \bigstrut[t]\\
	\hline
	SCC-LB  & 14   & 88   & 68   & 52 \bigstrut[t]\\
	\hline
\end{tabular}%
		\label{table5}%
	\end{minipage}

	\begin{minipage}{0.95\linewidth}
	\centering
	\footnotesize
	\caption{The Per-sequence Results on Different iOS Devices(ClassB-RA)}
\begin{tabular}{|p{4.5em}<{\centering}|c|c|c|c|c|c|}
	\hline
	\multicolumn{1}{|c|}{\multirow{3}[2]{*}{Sequence}} & \multirow{3}[2]{*}{QP} &      & VTM & \multicolumn{1}{p{2.5em}|}{O266 } & \multicolumn{1}{p{2.5em}|}{O266 } & \multicolumn{1}{p{2.5em}|}{O266 } \bigstrut[t]\\
	\multicolumn{1}{|c|}{} &      & \multicolumn{1}{p{2.5em}|}{Bitrate} &  \multicolumn{1}{p{2.5em}|}{Thread1 } & \multicolumn{1}{p{2.5em}|}{Thread1 } & \multicolumn{1}{p{2.5em}|}{Thread1 } & \multicolumn{1}{p{2.5em}|}{Thread1 } \\
	\multicolumn{1}{|c|}{} &      & \multicolumn{1}{p{2.44em}|}{ [Mbps]} & \multicolumn{1}{p{2.4em}|}{[fps] (A14)} & \multicolumn{1}{p{2.4em}|}{[fps] (A14)} & \multicolumn{1}{p{2.4em}|}{[fps] (A13)} & \multicolumn{1}{p{2.4em}|}{[fps] (A12)} \bigstrut[t]\\
	\hline
	\multicolumn{1}{|c|}{\multirow{4}[8]{*}{MarketPlace}} & 22   & 11.49 & 7    & 39   & 30   & 24 \bigstrut[t]\\
	\cline{2-7}\multicolumn{1}{|c|}{} & 27   & 4.68 & 8    & 45   & 35   & 28 \bigstrut[t]\\
	\cline{2-7}\multicolumn{1}{|c|}{} & 32   & 2.10  & 9    & 51   & 39   & 31 \bigstrut[t]\\
	\cline{2-7}\multicolumn{1}{|c|}{} & 37   & 0.96 & 12   & 61   & 47   & 37 \bigstrut[t]\\
	\hline
	\multirow{4}[8]{*}{RitualDance} & 22   & 8.96 & 7    & 42   & 33   & 26 \bigstrut[t]\\
	\cline{2-7}\multicolumn{1}{|c|}{} & 27   & 4.47 & 8    & 50   & 38   & 28 \bigstrut[t]\\
	\cline{2-7}\multicolumn{1}{|c|}{} & 32   & 2.34 & 10   & 58   & 44   & 36 \bigstrut[t]\\
	\cline{2-7}\multicolumn{1}{|c|}{} & 37   & 1.25 & 11   & 67   & 52   & 41 \bigstrut[t]\\
	\hline
	\multirow{4}[8]{*}{Cactus} & 22   & 8.98 & 8    & 45   & 35   & 26 \bigstrut\\
	\cline{2-7}\multicolumn{1}{|c|}{} & 27   & 3.50  & 9    & 57   & 44   & 34 \bigstrut[t]\\
	\cline{2-7}\multicolumn{1}{|c|}{} & 32   & 1.68 & 11   & 68   & 52   & 39 \bigstrut[t]\\
	\cline{2-7}\multicolumn{1}{|c|}{} & 37   & 0.85 & 13   & 74   & 58   & 52 \bigstrut[t]\\
	\hline
	\multicolumn{1}{|c|}{} & 22   & 10.4 & 7    & 40   & 30   & 23 \bigstrut[t]\\
	\cline{2-7}Basketball & 27   & 4.17 & 7    & 47   & 36   & 28 \bigstrut[t]\\
	\cline{2-7}Drive & 32   & 2.00    & 9    & 53   & 41   & 31 \bigstrut[t]\\
	\cline{2-7}\multicolumn{1}{|c|}{} & 37   & 1.03 & 10   & 53   & 41   & 35 \bigstrut[t]\\
	\hline
	\multirow{4}[8]{*}{BQTerrace} & 22   & 17.9 & 7    & 34   & 27   & 21 \bigstrut[t]\\
	\cline{2-7}\multicolumn{1}{|c|}{} & 27   & 3.49 & 9    & 51   & 39   & 31 \bigstrut[t]\\
	\cline{2-7}\multicolumn{1}{|c|}{} & 32   & 1.44 & 10   & 58   & 45   & 36 \bigstrut[t]\\
	\cline{2-7}\multicolumn{1}{|c|}{} & 37   & 0.70  & 12   & 66   & 50   & 41 \bigstrut[t]\\
	\hline
	Average & /    & /    & 9    & 53   & 41   & 32 \bigstrut[t]\\
	\hline
\end{tabular}%
	\label{table6}%
\end{minipage}
\end{table}

\section{Conclusion}\label{sec5}
In this paper, we present a high optimized H.266/VVC software decoder, which can support the real-time decoding on the ARM platform. On top of the framework of the published real-time HD/UHD decoder on the x86 platform, a uniform-designed software H.266/VVC decoder that can run real-time on different platforms and supports versatile functionalities is obtained. It has been shown that based on iPhone 12pro, 30fps decoding speed can be achieved for 4K bitstreams with SIMD acceleration and multi-threading processing.

\end{document}